\documentclass[prl,aps,twocolumn,twoside,showkeys]{revtex4-1}

\usepackage{amsmath,amsfonts,amssymb,amsthm,mathtools}

\usepackage[english]{babel}
\usepackage[utf8]{inputenc}
\usepackage[T1]{fontenc}

\usepackage{latexsym}
\usepackage[pdftex]{graphicx}
\usepackage{epstopdf}

\usepackage{subfigure}
\usepackage{MnSymbol}

\usepackage{enumitem}

\usepackage{hyperref}
\usepackage[all]{hypcap}

\hypersetup{
	pdftitle={},
	pdfauthor={},
	colorlinks=true,
	linkcolor=black,
	citecolor=black,
	filecolor=black,
	urlcolor=black,
	bookmarksopen,
	bookmarksopenlevel=1,
}

\newcommand{\cE}{\mathcal{E}}

\newcommand{\cJ}{\mathcal{J}}

\newcommand{\cO}{\mathcal{O}}

 \let\b=\beta         \let\d=\delta     
       \let\th=\theta    \let\ka=\kappa    \let\l=\lambda
                          
 \let\t=\tau            
   \let\o=\omega     
\let\G=\Gamma \let\D=\Delta

\newcommand{\ee}{\mathrm{e}}
\newcommand{\ii}{\mathrm{i}}

\def\io{\infty}

\def\tt{\tilde{t}}
\def\tx{\tilde{x}}

\newcommand{\ccS}{S}
\newcommand{\tE}{\tilde{E}}
\newcommand{\tJ}{\tilde{J}}

\renewcommand{\Re}{\operatorname{Re}}

\theoremstyle{definition}

\theoremstyle{remark}

\begin{document}


\title{%
Diffusive Heat Waves in Random Conformal Field Theory
}%

\author{Edwin Langmann}
\email{langmann@kth.se}
\affiliation{Department of Physics, KTH Royal Institute of Technology,
106 91 Stockholm, Sweden}

\author{Per Moosavi}
\email{pmoosavi@kth.se}
\affiliation{Department of Physics, KTH Royal Institute of Technology,
106 91 Stockholm, Sweden}

\date{%
February 14, 2019
}%

\begin{abstract}
We propose and study a conformal field theory (CFT) model with random position-dependent velocity that, as we argue, naturally emerges as an effective description of heat transport in one-dimensional quantum many-body systems with certain static random impurities.
We present exact analytical results that elucidate how purely ballistic heat waves in standard CFT can acquire normal and anomalous diffusive contributions due to our impurities.
Our results include impurity-averaged Green's functions describing the time evolution of the energy density and the heat current, and an explicit formula for the thermal conductivity that, in addition to a universal Drude peak, has a nontrivial real regular contribution that depends on details of the impurities.
\end{abstract}



\maketitle


\emph{Introduction.}---%
\csname phantomsection\endcsname
\addcontentsline{toc}{section}{Introduction}%
%
%
Heat transport has been modeled successfully by the diffusion equation since Fourier's time.
Still, the mathematical derivation of diffusion from microscopic models has remained an outstanding challenge \cite{BLRb}.
Recent efforts addressing this problem in one spatial dimension (1d) have led to important progress \cite{TTinLD}, including derivations in \cite{dNBD, IdNMP} of diffusive effects within hydrodynamical descriptions of integrable quantum many-body systems \cite{CBT, BCNF, BeDo3, Spo2}. 
Meanwhile, exact results from conformal field theory (CFT), routinely used to effectively describe universal properties of 1d quantum many-body systems, show that standard CFT only supports purely ballistic transport, see, e.g., \cite{CaCa1, BeDo1, BeDo2, GLM}. 
This points to the importance of randomness and impurities.
Such a route was recently explored in \cite{BeDo4} with positive results for CFT extended by special impurities that vary in time and space.
This, however, does not shed light on the questions if and how \emph{static} impurities in CFT can lead to diffusion.

In this Letter we propose and study CFT with random position-dependent velocity $v(x)$.
Such a random CFT, we argue, emerges naturally as an effective description of 1d quantum many-body systems with static random impurities that vary on mesoscopic length scales and are \emph{commensurate} in the sense that they induce the same spatial variations in all terms in the Hamiltonian.
Defining this velocity in terms of a Gaussian random function and using recent generalizations of CFT to inhomogeneous situations \cite{Kat2, WRL, DSVC, DSC, GLM}, we obtain exact analytical results which elucidate how ballistic transport in standard CFT can acquire normal and anomalous diffusive contributions due to such impurities. 

\begin{figure}[!htbp]

\centering
\includegraphics[scale=1, trim=10 7 4 20, clip=true]{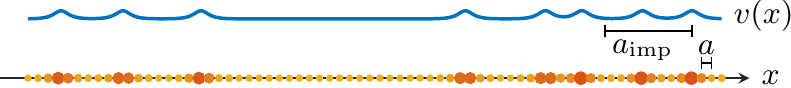}

\caption{%
Illustration of a velocity $v(x)$ effectively describing a lattice system with impurities varying on a mesoscopic scale $a_{\textnormal{imp}}$ much larger than the lattice spacing $a$.
For the $XXZ$ spin chain described in the main text, $x=ia$, and the color and size of the dots indicate the magnitude of the couplings $J_i$.%
}
\label{Fig:Random_inhomogeneous_CFT_subsystem}

\end{figure}

For example, consider a generalized $XXZ$ spin chain with uniformly varying couplings $J_i^x = J_i^y = J_i$ and $J_i^z=J_i \Delta$ between spins on adjacent sites $i$ and $i+1$, with constant $\Delta$.
As argued in \cite{DSVC, DSC}, if $J_i$ varies on length scales much larger than the lattice spacing and if $|\Delta|<1$ (gapless regime), then a generalized Luttinger model with position-dependent velocity $v(x)$ provides an effective description of this system.
Such a model is an example of inhomogeneous CFT with central charge $c=1$. 
We propose to use inhomogeneous CFT with random $v(x)$ to effectively describe, e.g., generalized $XXZ$ spin chains with couplings $J_i$ modeling static random impurities that are commensurate (constant $\D$) and vary on mesoscopic scales, see Fig.~\ref{Fig:Random_inhomogeneous_CFT_subsystem}.
The usual derivation in, e.g., \cite{SCP} is straightforward to generalize: the length-scale condition translates to sufficiently fast decay of $J_i$ in Fourier space, meaning only forward and no umklapp scattering is induced by our impurities, and, since they are commensurate, the only effect is that the constant velocity $v$ in the CFT description is replaced by $v(x)$ obtained from the $J_i$:s in the continuum limit \cite{Note1}.

We study heat transport in our random CFT model in two complementary ways that make use of established mathematical tools from wave propagation in random media \cite{Ish}.
Approach~A:  By deriving and exactly solving effective equations for heat transport.
Approach~B:  By computing the linear-response thermal conductivity as a function of frequency and deriving an explicit formula for its real regular part.
Both approaches are nonperturbative, and Approach~A is beyond linear response.

Averaging over impurities, we find heat waves that deform diffusively, and we obtain exact results for the thermal diffusivity $\alpha_{\textnormal{th}}$ at long times and the zero-frequency limit $L_{\textnormal{th}}$ of the real regular thermal conductivity.
Our results show that, in general, there are normal and anomalous diffusive contributions to heat transport on top of a ballistic one.
We also verify the Einstein relation between $\alpha_{\textnormal{th}}$ and  $L_{\textnormal{th}}$, which establishes a link between Approaches A and B.

In real systems, heat transport is ballistic at very low temperatures, but diffusion caused by randomness becomes increasingly important as temperature increases.
Our model allows one to analytically study the interplay between both kinds of transport.
To our knowledge, it is a new quantum model in condensed matter physics, which has similarities with models in geophysics and electrical engineering \cite{Ish}.
Importantly, the diffusion mechanism realized in our model seems to have been largely overlooked in the condensed matter literature.


\emph{Random CFT.}---%
\csname phantomsection\endcsname
\addcontentsline{toc}{section}{Random CFT}%
%
%
By \emph{inhomogeneous CFT} we mean a quantum field theory with Hamiltonian 
\begin{equation}
\label{Hamiltonian_inhomog_CFT}
H = \int dx\, v(x)[T_{+}(x) + T_{-}(x)],
\end{equation}
where $v(x)>0$ is a velocity that varies smoothly in space and $T_{\pm}(x)$ are operators satisfying commutation relations well known from standard CFT \cite{FMS}:%
\begin{multline}
\label{T_rx_T_rpxp}
[T_{\pm}(x), T_{\pm}(y)]
= \mp 2\ii\d'(x-y) T_{\pm}(y) \\
	\pm \ii\d(x-y)T_{\pm}'(y)
	\pm \frac{c}{24\pi}\ii\d'''(x-y)
\end{multline}
and $[T_{\pm}(x), T_{\mp}(y)] = 0$, with $c>0$ the central charge. 
The time evolution of observables $\cO$ is determined by the Heisenberg equation $\partial_t\cO = \ii[H, \cO]$ (we set $\hbar = k_B = 1$). 
In a specific model, $T_\pm(x)$ are represented by operators on a particular Fock space; there are many examples of interest, including the Luttinger model already mentioned, see, e.g., \cite{GLM}.
The special case $v(x)=v$ corresponds to standard CFT. 

We define \emph{random CFT} as inhomogeneous CFT with random velocity
\begin{equation}
\label{v_xi}
v(x) = v/[1-\xi(x)]
\end{equation}
with $\xi(x)$ a Gaussian random function \cite{Lif} specified by $\mathbb{E}[\xi(x)] = 0$ and the covariance
\begin{equation}
\label{G_xi}
\Gamma(x-y) = \mathbb{E}[\xi(x)\xi(y)],
\end{equation}
where $\mathbb{E}[\cdot]$ denotes the average over impurities. 
We assume that $\Gamma(x)$ is even, has nonnegative Fourier transform, and has finite $\Gamma_0 = \int dx\, \Gamma(x)$.
The parameter $\Gamma_0$ is nonnegative and has the dimension of length.
We find it convenient to introduce another length parameter $a_0>0$ to write $\Gamma(x) = (\Gamma_0/a_0)f(|x|/a_0)$, with $f(u)$ some suitable function of the dimensionless variable $u = |x|/a_0$.
Four illustrative examples, (a)--(d), of such functions $f(u)$ are given in Table~\ref{Table:Examples}.
Note that standard CFT can be recovered by setting $\Gamma_0 = 0$.

Equation \eqref{v_xi} can be understood as follows.
In an inhomogeneous system, the time for an excitation
to travel a fixed distance $dx$ is changed from $dx/v$ to $dx/v(x)$, and in our model, this time change is random.
We model this randomness by a Gaussian random function, and, as we will show, such random time changes lead to diffusion while preserving exact solvability.

We recall the following well-known property of Gaussian random functions:%
\begin{equation}
\label{main_property}
\mathbb{E} \bigl[ \ee^{-\ii \l \int_{y}^{x} \! d\tx\, \xi(\tx)} \bigr]
= \ee^{-\l^2 \Lambda(x-y)/2}
\end{equation}
for real $\lambda$ with $\Lambda(x-y) = \int_y^x dx_1 \int_y^x dx_2\, \Gamma(x_1-x_2)$.
This identity enables us to compute impurity averages in our model exactly.
Note that $\Lambda(x)$ is even, $\geq 0$, and 
$\Lambda(x) = \Gamma_0 a_0F(|x|/a_0)$
with 
$F(u) = \int_0^u dv_1 \int_0^u dv_2\, f(|v_1-v_2|)$;
the latter function is given in Table~\ref{Table:Examples} for Examples~(a)--(d).
As will be seen, certain transport properties will depend on details of the impurities, i.e., the specific form of the covariance $\G(x)$ in \eqref{G_xi}, and this dependence is encoded by the function $F(u)-u$.
There are also universal results independent of the latter function, which makes clear that Example~(a) is special in that it describes only the universal transport properties.

\renewcommand{\arraystretch}{1.5}
\begin{table}

\centering
\begin{tabular}{|c||c|c|c|c|}
	\hline
			& $f(u)$
			& $F(u)$
			& $F'(u)$ \\
	\hline \hline
	(a) & $\delta(u)$
			& $u$
			& $1$ \\
	\hline
	(b) & $\frac{1}{2}\ee^{-u}$
			& $u + \ee^{-u} - 1$
			& $1 - \ee^{-u}$ \\
	\hline
	(c) & $\frac{1}{2}\delta(u) + \frac{1}{4}\delta(1-u)$
			& $u + \frac{1-u}{2}\theta(1-u) - \frac{1}{2}$
			& $1 - \frac{1}{2}\theta(1-u)$ \\
	\hline
	(d) & $u/(1+u^2)^2$
			& $u - \arctan(u)$
			& $1-1/(1+u^2)$ \\
	\hline
\end{tabular}

\caption{%
Examples~(a)--(d) of functions $f(u)$ for $u \geq 0$ defining covariance functions as $\Gamma(x) = (\Gamma_0/a_0) f(|x|/a_0)$.
The functions $F(u)$ and $F'(u)$ are discussed in the text.
[$\delta(u)$ and $\theta(u)$ denote the Dirac delta and the Heaviside function.]%
}
\label{Table:Examples}

\end{table}

We note that the model with fixed impurities can be made mathematically precise using Minkowskian CFT on a circle.
As in \cite{GLM}, this model can be solved by straightening out $v(x)$ using conformal transformations and taking the thermodynamic limit.
However, we will mainly use simpler arguments to derive our results in the present Letter.


\emph{Approach~A: Heat waves in random media.}---%
\csname phantomsection\endcsname
\addcontentsline{toc}{section}{Approach A: Heat waves in random media}%
%
%
In standard CFT, the energy density and heat current operators are $\cE(x) = v[T_{+}(x)+T_{-}(x)]$ and $\cJ(x) = v^2[T_{+}(x)-T_{-}(x)]$, and their expectation values in an arbitrary state, $E(x,t) = \langle\cE(x,t)\rangle$ and $J(x,t) = \langle\cJ(x,t)\rangle$, satisfy
$\partial_t E(x,t) + \partial_x J(x,t) = 0$ and 
$\partial_t J(x,t) + v^2\partial_x E(x,t) = 0$, see, e.g., \cite{GLM}.
It is straightforward to generalize this to inhomogeneous CFT \cite{SM}:%
\begin{subequations}
\label{effective_eqs_of_motion}
\begin{gather}
\partial_t E(x,t) + \partial_x J(x,t) = 0, \\
\partial_t J(x,t) + v(x) \partial_x \bigl[ v(x) E(x,t) \bigr] = 0.
\end{gather}
\end{subequations}

We are interested in the situation where energy is injected into an equilibrium state at time $t=0$.
This corresponds to the initial conditions $E(x,0) = e_0(x)$ and $J(x,0) = 0$, where $e_0(x)$ describes an initial energy distribution.
This function $e_0(x)$ is arbitrary and independent of $v(x)$.

Inspired by \cite{KaKe, Bla}, we use tools from wave propagation in random media to compute $e(x,t) = \mathbb{E}[E(x,t)]$ and $j(x,t) = \mathbb{E}[J(x,t)]$.
Note that translation invariance, which is broken by the impurities, is recovered after averaging.
Our results can be written in terms of the impurity-averaged Green's functions
\begin{equation}
\label{G_r_xt}
G_{\pm}(x,t)
=	\th(\pm x)\frac{\ee^{-(x \mp vt)^2/2\Lambda(x)}}{\sqrt{2\pi \Lambda(x)}}
\end{equation}
as follows \cite{SM}:%
\begin{subequations}
\label{EE_EJ_xi}
\begin{align}
e(x,t)
& = \int dy\,
		\Bigl[
			G^{\cE}_{+}(x-y,t) + G^{\cE}_{-}(x-y,t)
		\Bigr] e_0(y),
	\label{EE_E_xi} \\
j(x,t)
& = \int dy\,
		\Bigl[
			G^{\cJ}_{+}\hspace{-0.66mm}(x-y,t) + G^{\cJ}_{-}\hspace{-0.66mm}(x-y,t)
		\Bigr] e_0(y),
\end{align}
\end{subequations}
with
\begin{subequations}
\label{G_EJ}
\begin{align}
G^{\cE}_\pm(x,t)
& = \frac{1}{2}
		\biggl( 1 - \frac{(x\mp vt)\Lambda'(x)}{2\Lambda(x)} \biggr)
		G_{\pm}(x,t), \\
G^{\cJ}_\pm(x,t)
& = \pm \frac{v}{2} G_{\pm}(x,t).
\end{align}
\end{subequations}

A few remarks are in order:
(i) Since $G_{\pm}(x,t)\to \d(x)$ as $t\to0^{+}$, the initial conditions are satisfied.
(ii) Since $G_{\pm}(x,t)\to \th(\pm x)\d(x\mp vt)$ as $\Gamma_0\to0$, the standard CFT results of \cite{LLMM2, GLM} are recovered.
(iii) Total energy is conserved in \eqref{EE_E_xi} \cite{SM}.

The functions in \eqref{G_r_xt} are Gaussian distributions with variance $\Lambda(x)$.
They provide an explicit description of how heat spreads in our system:  $G_\pm(x,t)$ describe waves moving to the right ($+$) or left ($-$) with speed $v$.
However, different from standard CFT, these heat waves are not purely ballistic: in general, as they move, their widths increase gradually, which indicates diffusion. 

To characterize this diffusive behavior we note that $G_{\pm}(x,t)$ solves the propagation-diffusion equation \cite{BGL}
\begin{equation}
\label{propagation-diffusion_eq}
\bigl[ v^{-1} \partial_t \pm \partial_x - \gamma(x) \partial_t^2 \bigr]
	G_{\pm}(x,t) = 0 
\end{equation}
for $\pm x > 0$ and $t > 0$, with
$\gamma(x)=\pm \Lambda'(x)/2v^2 = (\Gamma_0/2v^2) F'(|x|/a_0) > 0$
becoming constant for large $|x|$.
Phenomena described by a partial differential equation of the form in \eqref{propagation-diffusion_eq} are referred to as \emph{temporal diffusion} in \cite{BGL}, with $\gamma(x)$ a temporal diffusion coefficient \cite{Note2}.
This is similar to the usual notion of diffusion, the difference being that space and time have switched roles.

It is important to note that one can also interpret the above as standard diffusion in a frame of reference moving with a heat wave.
To see this, change variables to
$\tx = x \mp  vt$, $\tt = |x|/v$
and define
$\tilde{G}_{\pm}(\tx,\tt) = G_{\pm}(x,t)$.
This is a natural choice: $\tx$ is the coordinate of the observer moving with the wave, and $\tt$ is her time measured by the position of the wave. 
Equation \eqref{propagation-diffusion_eq} then becomes
\begin{equation}
\label{diffusion_eq}
\bigl[ \partial_{\tt} - \alpha_{\textnormal{th}}(\tt) \partial_{\tx}^2 \bigr]
	\tilde{G}_{\pm}(\tx,\tt) = 0 
\end{equation}
for $\tt > 0$ and $\pm\tx > -vt$, with the thermal diffusivity
$\alpha_{\textnormal{th}}(\tt) = (\Gamma_0v/2)F'(v\tt/a_0)$.
In general, $\alpha_{\textnormal{th}}(\tt)$ is time dependent, see Table~\ref{Table:Examples} for $F'(u)$ in our examples.
The exception is Example~(a), where it is equal to the constant
\begin{equation}
\label{alpha_th}
\alpha_{\textnormal{th}}
= \frac{\Gamma_0v}{2},
\end{equation}
while it converges to this value for large $v\tt/a_0$ in Examples~(b)--(d).

Equation \eqref{diffusion_eq} is a diffusion equation in a moving frame (the underlying ballistic motion) with heat waves changing according to a diffusion process given by
$\alpha_{\textnormal{th}}(\tt)$.
Equivalently, the variance of this process is $\Lambda(x)$, which in the new coordinates equals
$2\int_{0}^{\tt} dt'\, \alpha_{\textnormal{th}}(t') = \Gamma_0a_0 F(v\tt/a_0)$
and thus goes as $\Gamma_0 v \tt$ plus a nonlinear correction term, see Table~\ref{Table:Examples}.
This indicates that there are both normal and anomalous diffusive contributions \cite{LLP} on top of a ballistic one.
The normal diffusion is determined by the leading term $u$ of the function $F(u)$ and is in this sense universal.
The anomalous diffusive part is determined by the subleading term $F(u)-u$ and is thus nonuniversal.


\emph{Approach~B: Linear-response theory.}---%
\csname phantomsection\endcsname
\addcontentsline{toc}{section}{Approach B: Linear-response theory}%
%
%
We consider the linear-response thermal conductivity $\ka_{\textnormal{th}}(\o)$ averaged over impurities as a function of frequency $\o$ \cite{SM, Kubo}.
In general, its real part can be partitioned as $\Re \ka_{\textnormal{th}}(\o) = D_{\textnormal{th}} \pi \d(\o) + \Re \ka_{\textnormal{th}}^{\textnormal{reg}}(\o)$,
where 
$D_{\textnormal{th}}$
is the thermal Drude weight and
$\ka_{\textnormal{th}}^{\textnormal{reg}}(\o)$
is the regular part, see, e.g., \cite{GLM, Spo}.
A nonzero $D_{\textnormal{th}}$ corresponds to a ballistic contribution, while a nonzero
$\Re \ka_{\textnormal{th}}^{\textnormal{reg}}(\o)$ for $\o = 0$ ($\neq 0$)
corresponds to a normal (anomalous) diffusive contribution \cite{LLP}.

\begin{figure}[!htbp]

\centering
\includegraphics[scale=1, trim=0 0 0 0, clip=true]{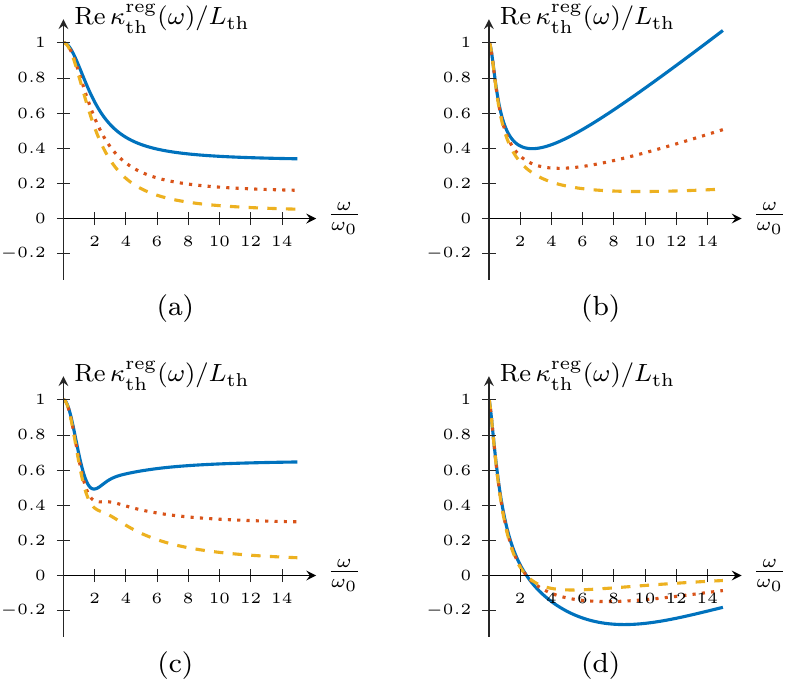}

\caption{%
$\Re \ka_{\textnormal{th}}^{\textnormal{reg}}(\o)$ in \eqref{EE_ka_f_th_reg} for Examples~(a)--(d) in Table~\ref{Table:Examples}.
In all plots, $\o_0 = v/\Gamma_0$, $\Gamma_0/a_0 = 0.6$ is fixed, and the parameter varied is $\beta\o_0$ equal to $1.8$ (blue solid line), $1.2$ (red dotted), and $0.6$ (yellow dashed).%
}
\label{Fig:Kappa_plots}

\end{figure}

For random CFT, our result is as described above with
$D_{\textnormal{th}}
= \pi vc/3\b$
and
\begin{align}
\Re \ka_{\textnormal{th}}^{\textnormal{reg}}(\o) 
& = \frac{\pi c}{6\b} 
	\biggl[ 1 + \left( \frac{\o\b}{2\pi} \right)^2 \biggr] \nonumber \\
& \;\;\;\, \times \!\int \!dx\,
	\ee^{-(1/2)(\o/v)^2  \Lambda(x)}
	\cos \biggl( \frac{\o x}{v} \biggr)
	\label{EE_ka_f_th_reg}
\end{align}
if $\Gamma_0 > 0$ and zero otherwise \cite{SM}.
The Drude weight is the same as in standard CFT, see, e.g., \cite{GLM}.
This corresponds to the well-known universality of ballistic heat transport in CFT, which extends to our situation with impurities. 
In addition, we obtain a nontrivial diffusive contribution described by
$\Re \ka_{\textnormal{th}}^{\textnormal{reg}}(\o)$ in \eqref{EE_ka_f_th_reg}, which is plotted for different $\Lambda(x)$ in Fig.~\ref{Fig:Kappa_plots}.

For $\Lambda(x) = \Gamma_0 |x|$ [our Example~(a)] one can compute the integral in \eqref{EE_ka_f_th_reg} analytically to obtain
$\Re \ka_{\textnormal{th}}^{\textnormal{reg}}(\o)
= (\pi c/6\b)
\bigl[ 1 + (\o\b/2\pi)^2 \bigr] \Gamma_0 \big/ \bigl[ 1 + (\o\Gamma_0/2v)^2 \bigr]$, which implies 
\begin{equation}
\label{L_th}
L_{\textnormal{th}}
= \lim_{\o\to0} \Re \ka_{\textnormal{th}}^{\textnormal{reg}}(\o)
= \frac{\pi c}{6\b} \Gamma_0. 
\end{equation}
This is actually true independent of details of the impurities \cite{SM}, and thus, in particular, also for Examples~(b)--(d).
Since $L_{\textnormal{th}}$ characterizes normal diffusion, this confirms that the normal diffusion in our model is universal.
Moreover, \eqref{alpha_th} and \eqref{L_th} imply $L_{\textnormal{th}} = (\pi c/3\b v) \alpha_{\textnormal{th}}$, which provides a link between our two approaches.
Since the volume specific heat capacity can be shown to be $c_V = \pi c/3 \beta v$ in both random and standard CFT, this verifies the Einstein relation $L_{\textnormal{th}} = c_{V} \alpha_{\textnormal{th}}$ for heat transport.

The behavior of $\Re \ka_{\textnormal{th}}^{\textnormal{reg}}(\o)$ for $\o \neq 0$ depends on impurity details, see Fig.~\ref{Fig:Kappa_plots}.
In particular, while $\Re \ka_{\textnormal{th}}^{\textnormal{reg}}(\o)$
becomes constant for large $\o$ in Examples~(a),~(c), and~(d) in Table~\ref{Table:Examples}, it grows linearly in Example~(b), and it can be seen to grow sublinearly in the example
$f(u) = \ee^{-\sqrt{u}}/4\sqrt{u}$.
It would be interesting to explore this dependence on details more systematically.

%


\emph{Conclusions.}---%
\csname phantomsection\endcsname
\addcontentsline{toc}{section}{Conclusions}%
%
%
We proposed and studied an exactly solvable CFT model with random position-dependent velocity.
Such a model, we argued, naturally emerges as an effective description of 1d quantum many-body systems with commensurate static random impurities varying on mesoscopic length scales.
We presented two exact results for heat transport that prove, in complementary ways, that such impurities can lead to diffusive contributions on top of the well-known ballistic one of standard CFT.
In particular, we found a universal normal diffusive contribution and, in general, nonuniversal anomalous diffusive contributions that depend on details of the impurities.

Our impurities do not lead to umklapp or backward scattering.
This is different from more common impurity models, but those are also generally more difficult to treat.
Our model shows that impurities leading to only forward scattering can be included in CFT without spoiling the exact solvability, and that they can lead to diffusion.
It is important to investigate if this captures universal features of real systems.
However, there are good reasons to be optimistic: different scattering mechanisms often manifest themselves in the same way on larger scales, e.g., heat transport in different systems is modeled successfully by a Boltzmann equation with the same approximate collision term \cite{BGK}.

The diffusion mechanism in our model has a simple interpretation as follows.
For fixed impurity configuration, consider a source at position $x_0$ emitting a deltalike heat pulse at time $t_0$ and monitor its arrival time $t_1$ at position $\pm x_1 > 0$.
If $v(x)$ and the positions $x_0$ and $x_1$ were known accurately on microscopic scales, one could compute $t=t_1-t_0$ deterministically. 
However,  we are interested in situations where $x_0$ and $x_1$ are macroscopic lengths, i.e., only known to accuracies much larger than the average impurity distance. 
The region between $x_0$ and $x_1$ thus has to be modeled as a random medium, and the propagation time $t$ is given by a probability distribution $G_\pm(x,t)$ with $x=x_1-x_0$. By general probabilistic arguments, one expects that this distribution satisfies a partial differential equation as in \eqref{propagation-diffusion_eq} \cite{BGL}, which our results confirm.

Previous works deriving diffusion from static impurities in microscopic models using precise mathematical arguments include \cite{LuSp, BaOl, BHLLO} for classical systems and \cite{ESY} for a noninteracting quantum system.
As far as we know, there are no such previous rigorous results for interacting quantum systems.
We thus stress that our exact results are for quantum systems that include interacting ones.

We finally note that models similar to the generalized $XXZ$ spin chain discussed in the introduction have received a lot of attention in the context of many-body localization \cite{BPSS}.
It would be interesting to investigate if potential signatures of many-body localization can be established in random CFT, similarly as in \cite{BeDo4}.


\begin{acknowledgments}
We are very grateful to Krzysztof Gaw\k{e}dzki for collaboration at an early stage of this work.
We would also like to thank Jens Bardarson, Benjamin Doyon, Joel Lebowitz, Vieri Mastropietro, Wojciech De Roeck, Herbert Spohn, and Mats Wallin for helpful discussions. 
E.L.\ acknowledges support by the Swedish Research Council (Grant No.\ 2016-05167).
P.M.\ is thankful for financial support from the Knut and Alice Wallenberg Foundation (No.\ V-2018-0505) and ``Roland Gustafssons stiftelse f{\"o}r teoretisk fysik.''
\end{acknowledgments}




\onecolumngrid
\clearpage


\section{Supplemental Material}


Part~A contains computational details for Approach~A [Eqs.~\eqref{effective_eqs_of_motion}--\eqref{G_EJ}] and Part~B for Approach~B [Eqs.~\eqref{EE_ka_f_th_reg} and~\eqref{L_th}].


\bigskip

\begin{center}
\bfseries\small Part~A: Heat Waves in Random Media
\end{center}

\smallskip


The energy density operator given by \eqref{Hamiltonian_inhomog_CFT} and the corresponding heat current operator are
\begin{equation}
\cE(x)=v(x)[T_+(x)+T_-(x)],
\quad
\cJ(x) = v(x)^2 [T_{+}(x) - T_{-}(x)].
\end{equation}
Using the Heisenberg equation and \eqref{T_rx_T_rpxp}, one verifies that
\begin{equation}
\partial_{t} \cE(x,t) + \partial_{x} \cJ(x,t) = 0,
\quad
\partial_t \cJ(x,t) + v(x) \partial_x \bigl[ v(x)\cE(x,t) + \ccS(x) \bigr] = 0,
\end{equation}
where
$\ccS(x) = -(c/12 \pi) \bigl[ v(x)v''(x) - v'(x)^2/2 \bigr]$
is an anomaly originating from the Schwinger term in \eqref{T_rx_T_rpxp}.
These equations of motion guarantee conservation of energy but not of the total current $\int dx\, \cJ(x)$ or the momentum $\int dx\, [T_+(x)-T_-(x)]$.

We define $\tE(x,t) = \langle \cE(x,t) \rangle$ and $\tJ(x,t) = \langle \cJ(x,t) \rangle$ with $\langle \cdot \rangle$ the expectation value in some arbitrary state in the thermodynamic limit.
The expectations satisfy the same equations of motion as the operators. 
They have the following static solutions: $\tE_{\textnormal{stat}}(x) = [C_1 - \ccS(x)]/v(x)$ and $\tJ_{\textnormal{stat}}(x) = C_2$ with real constants $C_{1}$ and $C_{2}$, which describe an equilibrium state if $C_2=0$ and a nonequilibrium steady state if $C_2 \neq 0$.
It follows that $E(x,t) = \tE(x,t) - \tE_{\textnormal{stat}}(x)$ and $J(x,t) = \tJ(x,t) - \tJ_{\textnormal{stat}}(x)$ satisfy \eqref{effective_eqs_of_motion}.

To solve \eqref{effective_eqs_of_motion} with our initial conditions, we observe that we can write
\begin{equation}
E(x,t) = [{u_{+}(x,t)} + {u_{-}(x,t)}]/{v(x)},
\quad
J(x,t) = {u_{+}(x,t)} - {u_{-}(x,t)},
\end{equation}
with  $u_{\pm}(x,t)$ satisfying 
$\partial_t u_{\pm}(x,t)	\pm v(x) \partial_x u_{\pm}(x,t) = 0$.
Our initial conditions $E(x,0) = e_0(x)$ and $J(x,0) = 0$ translate into
$u_{\pm}(x,0) = v(x) e_0(x)/2$. 
Using standard methods for partial differential equations, we find the following exact solution for the initial value problem for $u_{\pm}$:%
\begin{equation}
u_{\pm}(x,t)
= \int \!dy\, \frac{\th(\pm(x-y))}{2}
	\int \frac{d\o}{2\pi}
	\ee^{\ii\o \int_{y}^{x} \!d\tx\, v(\tx)^{-1} \mp \ii\o t} e_0(y)
\end{equation}
with the Heaviside function $\theta(x)$. 
Inserting \eqref{v_xi}, using \eqref{main_property} to compute the impurity average, and using a standard Gaussian integral, we obtain
\begin{equation}
\mathbb{E}[u_{\pm}(x,t)]
= \int dy\, \frac{v}{2} G_{\pm}(x-y,t) e_0(y)
\end{equation}
with $G_{\pm}(x,t)$ in \eqref{G_r_xt}.
In a similar manner we compute $\mathbb{E}[u_{\pm}(x,t)/v(x)]$.
From this, the results in \eqref{EE_EJ_xi} and \eqref{G_EJ} follow.

Energy conservation in \eqref{EE_E_xi} can be shown as follows.
We can write 
$G^{\cE}_\pm(x,t) = [\th(\pm x)/2] (2\pi)^{-1/2} \partial_{x} [\chi_{\pm}]
\ee^{-\chi_{\pm}^2/2}$
with
$\chi_{\pm} = (x\mp vt)/\sqrt{\Lambda(x)}$.
Using this, one finds 
$\int \!dx\, e(x,t) = \int \!dx\, e_0(x)$
by a change of variables to $\chi_{\pm}$ and computing a Gaussian integral.


\bigskip

\begin{center}
\bfseries\small Part~B: Linear-Response Theory
\end{center}

\smallskip


Let $\ka_{\textnormal{th},\xi}(\o)$ be the thermal conductivity at fixed impurity configuration indicated by the subscript $\xi$.
We define it as the response function related to the total heat current obtained by perturbing the equilibrium state at temperature $\b^{-1}$ with a unit pulse perturbation $V = -(\d\b/\b) \int dx\, W(x) \cE(x)$ at time zero, where $W(x)$ is a smooth function equal to $1/2$ ($-1/2$) to the far left (right), cf.\ [\ref{GLM_SM}, \ref{LLMM2_SM}]. 
Using standard linear-response theory [\ref{Kubo_SM}], one derives the Green-Kubo formula
\begin{equation}
\label{GK_1}
\ka_{\textnormal{th},\xi}(\o)
=	\b \!\int_{0}^{\b} \!d\t \!\int_{0}^{\io} \!dt\, 
	\ee^{\ii\o t}
	\!\int\!dx \int\! dx'\,
	\partial_{x'} \bigl[ -W(x') \bigr] 
	\bigl\langle \cJ(x,t) \cJ(x',i\t) \bigr\rangle_{\b}^{c},
\end{equation}
where $\langle \cJ(x,t) \cJ(x',\ii\t) \rangle_{\b}^{c}$ is the connected current-current correlation function in thermal equilibrium with respect to $H$ in
\eqref{Hamiltonian_inhomog_CFT}.
Since translational invariance is broken, we cannot change variables to do the $x'$-integral.

Using CFT results developed in [\ref{GLM_SM}], we derive an explicit formula for the correlation function in \eqref{GK_1}.
Computing the time integrals exactly using the residue theorem, we obtain $D_{\textnormal{th},\xi}
= \pi vc/3\b
= D_{\textnormal{th}}$ independent of $\xi$ and
\begin{equation}
\label{ka_f_th_reg}
\Re \ka_{\textnormal{th},\xi}^{\textnormal{reg}}(\o)
= \frac{\pi c}{6\b} 
	\biggl[ 1 + \left( \frac{\o\b}{2\pi} \right)^2 \biggr] 
	\!\int \!dx\int\! dx'\,
	\partial_{x'} \bigl[ -W(x') \bigr]
	\biggl( 1 - \frac{v}{v(x)} \biggr)
	\cos \biggl( \o \!\int_{x'}^{x} \frac{d\tx}{v(\tx)} \biggr).
\end{equation}
For standard CFT, \eqref{ka_f_th_reg} is zero.
To compute $\ka_{\textnormal{th}}(\o) = \mathbb{E}[\ka_{\textnormal{th},\xi}(\o)]$, we write the cosine as sum of exponentials, insert \eqref{v_xi}, and use \eqref{main_property}. 
After averaging, translation invariance is recovered, which allows us to do the $x'$-integral and obtain the result independent of $W(x)$ given in \eqref{EE_ka_f_th_reg}.

Lastly, that $L_{\mathrm{th}}$ is given by \eqref{L_th} independent of impurity details can be shown as follows.
We change the integration variable in \eqref{EE_ka_f_th_reg} to $\zeta = \o x/v$ and note that the function in the exponential becomes
$-(1/2) \Gamma_0 a_0 (\o/v)^2 F(v|\zeta|/\o a_0)$,
which equals
$-(1/2) \Gamma_0 \o |\zeta|/v$
up to subleading terms not contributing to the integral as $\o\to0$.
This implies the result in \eqref{L_th}.


\bigskip

\noindent {\bf References}
{\small%
\begin{enumerate}[leftmargin=2.0em, itemsep=0.0em, label={[\arabic*]}, ref={\arabic*}, itemindent=0.0em]

\setcounter{enumi}{11}

\item
\label{GLM_SM}
K.~Gaw\k{e}dzki, E.~Langmann, and P.~Moosavi,
``Finite-time universality in nonequilibrium CFT,''
J.\ Stat.\ Phys.\ {\bf 172}, 353 (2018).

\setcounter{enumi}{25}

\item
\label{LLMM2_SM}
E.~Langmann, J.~L.~Lebowitz, V.~Mastropietro, and P.~Moosavi,
``Time evolution of the Luttinger model with nonuniform temperature profile,''
Phys.\ Rev.\ B {\bf 95}, 235142 (2017).

\setcounter{enumi}{29}

\item
\label{Kubo_SM}
R.~Kubo,
``Statistical-mechanical theory of irreversible processes. I. General theory and simple applications to magnetic and conduction problems,''
J.\ Phys.\ Soc.\ Jpn.\ {\bf 12}, 570 (1957).

\end{enumerate}
}



\end{document}